\magnification = \magstep1
\pageno=1
\hsize=15.0truecm
\hoffset=1.0truecm
\vsize=23.5truecm
\pretolerance=3000
\tolerance=5000
\hyphenpenalty=10000    
\def\cl{\centerline}
\def\ni{\noindent}
\def\vs{\vskip 11pt}

\vs\vs\vs
\cl{\bf A CORRECTION TO THE PP REACTION}
\vs\vs
\cl{Robert L. Kurucz}
\vs
\cl{Harvard-Smithsonian Center for Astrophysics, 60 Garden St, 
Cambridge, MA 02138}
\vs\vs\vs 

\cl{ABSTRACT}\vs

     These descriptive comments are made to encourage detailed three-body, 
relativistic, quantum collision calculations for the pp reaction.

  In stars, coulomb barrier tunneling, as in the pp reaction, is not a 
two-body process.  Tunneling is mediated by an energetic electron that 
interacts with the colliding particles.  The presence of such an electron 
lowers the potential barrier and increases the probability of tunneling by 
orders of magnitude.  The solar luminosity can be maintained with a central 
temperature near 10$^{7}$K where the neutrino production rates correspond 
to the observed rates.  Current stellar interior and evolutionary models 
need substantial revision.
\vs
\ni Subject headings: neutrinos --- nuclear reactions --- stars: interiors 
--- sun: interior  
\vs
\vs

     As the pp reaction   p + p $>$ d + e$^{+}$ + $\nu$  can be produced 
in laboratory accelerators, there is no question that it is a real reaction 
in which two protons move toward each other with enough energy to 
quantum-mechanically tunnel through their repulsive coulomb potential and 
combine.  The proton-proton reaction, coulomb-barrier tunneling, and 
statistical Debye-H\"{u}ckel electron shielding are discussed in basic texts.  
However, the conditions in a dense plasma at the center of a star 
substantially differ from those in an accelerator.  At the center of the sun
the temperature is of the order of 15 $\times$ 10$^{6}$K, the proton density 
is of the order of 10$^{26}$ protons-cm$^{-3}$, and the electron density in 
of the order of 10$^{26}$ electrons-cm$^{-3}$.  Typical velocities are 500 
km s$^{-1}$ for protons and 20,000 km s$^{-1}$ for electrons.  Slowly moving 
electrons tend to cluster around slowly moving protons.  The electron cluster 
reduces the proton effective charge by a small amount near the proton and 
cuts it off completely at a radius of about 10$^{-9}$ cm.  Neither fast 
electrons nor fast protons are aware of a slow proton until they penetrate 
the shielding electron cluster at which point they are immediately attracted 
or repelled by the coulomb potential.  As the electrons typically move 40 
times faster than the protons, the electron-proton collision frequency must 
be about 40 times the proton-proton collision frequency.  A colliding fast 
proton decelerates from 2000 or 3000 km s$^{-1}$ to 0 km s$^{-1}$ relative 
velocity by the time it reaches a separation of 10$^{-10}$ cm, which is only 
90\% of the distance to the target proton.  Unless they tunnel, protons are 
always far apart on a nuclear scale because the nuclear interaction radius 
is on the order of 10$^{-13}$ cm.  A proton-proton collision is a slow   
\vfill
\eject
\ni
process.  An electron-proton collision is much faster.  A colliding fast 
electron passes through the shielding electron cluster at, say, 100,000 
km s$^{-1}$ and is immediately accelerated toward the central proton.  
In some collisions the electron passes near the proton, through the volume 
inaccessible in a proton-proton collision.  

     A proton can suffer both a proton and an electron collision 
simultaneously.  Such collisions may be infrequent, but they are more 
probable than tunneling, and they determine the pp reaction rate.  When a 
fast electron penetrates the electron cluster during a proton-proton 
collision it is attracted by both protons.  If the electron approaches 
from a polar direction with respect to the proton-proton axis, it helps 
to pull the protons apart and it prevents tunneling.  If the electron 
approaches equatorially, it shields the protons from each other and 
accelerates them toward each other.  Part of the kinetic energy of the 
electron contributes to the pp reaction.  When the electron leaves, the 
two protons are closer than they would have been on their own and the 
tunneling probability has greatly increased.  The reaction
p + p + e $>$ d + e + e$^{+}$ + $\nu$ requires lower proton energies than 
the reaction p + p $>$ d + e$^{+}$ + $\nu$.  A solar central temperature 
of, say, 10 $\times$ 10$^{6}$K produces the same energy and neutrino yield 
as 15 $\times$ 10$^{6}$K for the pp reaction without the electron boost.  
At 10 $\times$ 10$^{6}$K pp side chain reactions are much slower than at 
15 $\times$ 10$^{6}$K and produce the low neutrino rates that are actually 
observed.

     These descriptive comments are made to encourage detailed three-body, 
relativistic, quantum collision calculations for the pp reaction.  Until
such calculations become available, the problem can be investigated with
solar evolutionary models by making ad hoc increases in the pp reaction 
rate until the model yields the observed neutrino flux.  

      Beyond the pp reaction there is much more work.  
The reactions d + p, $^{3}$He + $^{3}$He, $^{3}$He + $^{4}$He, $^{7}$Be + p, 
and $^{7}$Li + p are coulomb barrier reactions and also have to be recalculated.
The Be and B neutrinos do not come from coulomb barrier reactions so they 
are not directly affected.

Light element burning occurs at lower temperatures than have been assumed.

\vfill\end